\documentclass{article}
\usepackage{amsmath}
\usepackage{amsfonts}
\usepackage{amssymb}
\usepackage{hyperref}

\begin{document}

\title{\textbf{Quantum Twist-Deformed }$D=4$\textbf{\ Phase Spaces with Spin
Sector and Hopf Algebroid Structures}}
\author{\textbf{Jerzy Lukierski}$^{1}$, \textbf{Stjepan Meljanac}$^{2}$, 
\textbf{Mariusz Woronowicz}$^{1}$ \vspace{12pt} \\
$^1$ Institute for Theoretical Physics, University of Wroclaw,\\
pl. Maxa Borna 9, 50-205 Wroclaw, Poland \\
$^{2}$Division of Theoretical Physics, Rudjer Bo\v{s}kovi\'{c} Institute, \\
Bijeni\v{c}ka c. 54, HR-10002 Zagreb, Croatia}
\date{}
\maketitle

\begin{abstract}
We consider the generalized $(10+10)-$dimensional $D=4$ quantum phase spaces
containing translational and Lorentz spin sectors associated with the dual
pair of twist-quantized Poincare Hopf algebra $\mathbb{H}$ and quantum
Poincare Hopf group $\widehat{\mathbb{G}}$. Two Hopf algebroid structures of
generalized phase spaces with spin sector will be investigated: first one $%
\mathcal{H}^{(10,10)}$ describing dynamics on quantum group algebra $%
\widehat{\mathbb{G}}$ provides by the Heisenberg double algebra $\mathcal{HD=%
}\mathbb{H}\rtimes \widehat{\mathbb{G}}$, and second, denoted by $\mathcal{%
\tilde{H}}^{(10,10)}$, describing twisted Hopf algebroid with base space
containing twisted noncommutative Minkowski space $\hat{x}_{\mu }$. We
obtain the first explicit example of Hopf algebroid structure of
relativistic quantum phase space which contains quantum-deformed Lorentz
spin sector.
\end{abstract}

\section{Introduction}

\bigskip Following recent description of noncommutative quantum phase spaces
as bialgebroids and Hopf algebroids (see \cite{lu}-\cite{lmmw}) we shall
study in this paper such algebraic structures in quantum phase spaces
derived from twisted quantum Poincare symmetries. In comparison with recent
studies \cite{IJMPA-twist}-\cite{lmmw} the novelty in our approach is the
appearance of Hopf algebroid description of $D=4$ generalized relativistic
quantum-deformed phase spaces with additional coordinates and momenta
describing Lorentz spin sector.

The quantum-deformed relativistic phase spaces $\mathcal{H}^{(4,4)}$,
spanned by the degrees of freedom $(\hat{x}_{\mu },\hat{p}_{\mu })$, are
described by quantum deformations of canonical relativistic Heisenberg
algebra\footnote{%
We denote the undeformed canonical phase space variables and canonical
Poincare algebra generators by letters without hats.}

\begin{equation}
\left[ p_{\mu },x_{\nu }\right] =-i\eta _{\mu \nu },\qquad \left[ x_{\mu
},x_{\nu }\right] =\left[ p_{\mu },p_{\nu }\right] =0.  \label{pf}
\end{equation}%
Such choice of phase space is suitable only for the description of standard
spinless dynamics. In this paper we shall consider the generalized $D=4$
quantum phase spaces $\mathcal{H}^{(10,10)}$ and $\mathcal{\tilde{H}}%
^{(10,10)}$. The additional generators $\hat{\Lambda}_{\mu \nu }$ will be
given by quantum counterpart of six Lorentzian angles $(\hat{\Lambda}\hat{%
\Lambda}^{T}=\hat{\Lambda}^{T}\hat{\Lambda}=I)$ dual to the generators $%
M_{\mu \nu }$ describing quantum-deformed Lorentz algebra.

The phase space description of spin dynamics as Heisenberg double has its
roots in the half century old idea of \ Souriau \cite{sourio} and Kostant 
\cite{kost} who described symplectic dynamics of point-like objects by the
geometry of cosets $\mathcal{K=}G/S$, where $G$ is the space-time symmetry
group, and $S$ its so-called stability group. If $G$ is the Poincare group $%
\mathcal{P}^{4}$, for providing the dynamics of spin degrees of freedom \
the coset $\mathcal{K}$ should include besides translations as well some
Lorentz group parameters (see e.g. \cite{bacry2}-\cite{bal2}). In such a way
one can describe e.g. the infinite-dimensional spin multiplets by adding to
coordinate sector the spinorial Weyl spinor coordinates $\eta _{\alpha
},\eta _{\dot{\alpha}}=\bar{\eta}_{\alpha }(\alpha =1,2)$ defined by
fourdimensional coset of $SL(2;\mathbb{C})$. In this way one can introduce
the generalized wave functions (classical fields) $\Psi (x;\eta _{\alpha
},\eta _{\dot{\alpha}})$ obtained if $S$ is a suitable two-dimensional
subgroup of Lorentz group\footnote{%
It can be shown ( see e.g. \cite{fink},\cite{bacry}) that for $D=4$ the
generalized field $\psi (x;\eta ,\bar{\eta})$ can be defined as introduced
on $8-$dimensional coset $\mathcal{P}^{4}/H_{2}$, where $H_{2}$ is generated
by Lorentz generators $(M_{01}+M_{31},M_{02}+M_{32})$ and $SL(2;\mathbb{C}%
)/H_{2}$ is parametrized by a pair $(\eta _{\alpha },\bar{\eta}_{\dot{\alpha}%
})$ of complex-conjugated $D=4$\ Weyl spinors. The idea of studying dynamics
on $10-$dimensional Poincare group manifold was early advocated by Lurcat 
\cite{lurc2}, \cite{lur}. Other way of introducing additional variables
describing spin is to supplement space-time by two-parameter auxiliary
manifold described by the sphere $S^{2}$ \cite{wigner2}-\cite{kuz}.}.

The Souriau-Kostant approach as well as its generalization to Poisson
manifolds \cite{kir},\cite{arno} and Poison-Lie theory \cite{drinff}-\cite%
{yu} can be also extended to models based on noncommutative (NC) geometry.
The construction of quantum-deformed NC\ space-times extended by the
additional quantum Lorentz group parameters has been recently also proposed
in the case of quantum-deformed Poincare group (see e.g. \cite{bale1}-\cite%
{bale2} where the case $D=2+1$ was considered).

In this paper we shall study firstly the generalized quantum-deformed phase
space $\mathcal{H}^{(10,10)}=(\hat{\xi}_{\mu },\hat{\Lambda}_{\mu \nu
};p_{\mu },M_{\mu \nu })$ with noncommutative generalized coordinates
described by the algebraic manifold of full quantum $D=4$ Poincare group
(i.e. $S=1$), and supplemented with dual algebra of generalized momenta
defined by ten generators of quantum-deformed Poincare algebra. The
quantization will be obtained by the twisting procedure considered recently
in \cite{lmmw} (see also \cite{dlw06}), which generate the Lie-algebraic
deformation of space-time translations algebra and the quadratic algebra
describing the commutators of all ten quantum Poincare group generators $(%
\hat{\xi}_{\mu },\hat{\Lambda}_{\mu \nu })$. Subsequently, following \cite%
{lu} we anticipate the Hopf algebroid structure describing the dynamics on
algebraic quantum Poincare group manifold.

Further, following \cite{xu},\cite{bp} we introduce $(10+10)$-dimensional
quantum phase space $\mathcal{\tilde{H}}^{(10,10)}=(\hat{x}_{\mu },\tilde{%
\Lambda}_{\mu \nu };p_{\mu },M_{\mu \nu })$ with Hopf algebroid structure
generated by twist-deformed $D=4$ Poincare-Hopf algebra $\mathbb{H}$, with
base space described by $\mathbb{H}$-module $\hat{X}_{A}=(\hat{x}_{\mu },%
\tilde{\Lambda}_{\mu \nu }),$ $(A=1\ldots 10)$, its noncommutativity
structure determined by twist-deformed star product multiplication%
\begin{equation}
\hat{X}_{A}\cdot \hat{X}_{B}\simeq X_{A}\star _{\mathcal{F}}X_{B}=m[\mathcal{%
F}^{-1}\circ (X_{A}\otimes X_{B})]=(\mathcal{F}_{(1)}^{-1}\rhd X_{A})(%
\mathcal{F}_{(2)}^{-1}\rhd X_{B})  \label{star}
\end{equation}%
where the Drinfeld twist factor $\mathcal{F}=\sum \mathcal{F}_{(1)}\otimes 
\mathcal{F}_{(2)}\in \mathbb{H\otimes H}$ satisfies the two-cocycle
condition \cite{xu},\cite{majid}. In nondeformed relativistic theory one can
identify the translations fourvectors $\hat{\xi}_{\mu }$ and Minkowski
space-time coordinates $\hat{x}_{\mu }$; in the presence of quantum
deformations the algebraic properties of base spaces $(\hat{\xi}_{\mu },\hat{%
\Lambda}_{\mu \nu })$ and $(\hat{x}_{\mu },\tilde{\Lambda}_{\mu \nu })$ are
usually different\footnote{%
In the canonical case of $\theta _{\mu \nu }$-defomation this difference is
exposed e.g. in \cite{dlw06}, \cite{koss}.}.

We shall consider therefore two possible types of Hopf algebroids describing
quantum phase spaces which are defined by twist-quantized Hopf algebra $%
\mathbb{H}$ (quantum deformed Poincare algebra) defining generalized momenta
and generalized coordinates, given by two choices of the $\mathbb{H}$-module:

\begin{enumerate}
\item[i)] described by Heisenberg double $\mathcal{H}\mathcal{D}=\mathbb{H}%
\rtimes \widehat{\mathbb{G}}$ defined by smash product\footnote{%
Smash product algebra (see e.g. \cite{bp2}) is a special kind of
cross-product algebra $\mathbb{H}\rtimes \mathbb{V}$ when the left $\mathbb{H%
}$-module $\mathbb{V}$ is provided by dual Hopf algebra $\widehat{\mathbb{G}}
$, with the action $\mathbb{H}\otimes \widehat{\mathbb{G}}\mathbb{%
\rightarrow }\widehat{\mathbb{G}}$ defined with the help of bilinear pairing 
$\langle \cdot ,\cdot \rangle :\mathbb{H}\otimes \widehat{\mathbb{G}}\mathbb{%
\rightarrow C}$. In the derivation of quantum-mechanical Heisenberg algebra
(see (\ref{pf})) as Heisenberg double the $\mathbb{C}$-number pairing is
assumed to be proportional to $\hbar $; in this paper we shall put $\hbar =1$%
.} of dual Hopf algebras $\mathbb{H},\widehat{\mathbb{G}}$, with built-in
Hopf action $h\blacktriangleright \hat{g}=\hat{g}_{(1)}\langle h,\hat{g}%
_{(2)}\rangle $ and $h\blacktriangleright (\hat{g}\hat{g}^{\prime
})=(h_{(1)}\blacktriangleright \hat{g})(h_{(2)}\blacktriangleright \hat{g}%
^{\prime })$ $(h\in $ $\mathbb{H};\hat{g},\hat{g}^{\prime }\in \widehat{%
\mathbb{G}};\Delta (h)=h_{(1)}\otimes h_{(2)})$. The associative
multiplication formula in $\widehat{\mathbb{G}}\otimes \mathbb{H}$ algebra
is given by%
\begin{equation}
(\hat{g}\otimes h)(\hat{g}^{\prime }\otimes h^{\prime })=\hat{g}%
(h_{(1)}\blacktriangleright \hat{g}^{\prime })\otimes h_{(2)}h^{\prime }
\label{smash}
\end{equation}%
where $\widehat{\mathbb{G}}$\ describes generalized coordinates.

Heisenberg double data are specified by the pair of dual Hopf algebras ($%
\mathbb{H},\widehat{\mathbb{G}}$) with the pairing $\langle \cdot ,\cdot
\rangle $ (see also $^{4)}$), which expresses the duality of $\mathbb{H}$
and $\widehat{\mathbb{G}}$ by the formulae%
\begin{eqnarray}
\langle h,\,\hat{g}\hat{g}^{\prime }\,\rangle \, &=&\langle \Delta (h),\,%
\hat{g}\otimes \hat{g}^{\prime }\,\rangle \,  \label{du1} \\
\langle hh^{\prime },\,\hat{g}\,\rangle \, &=&\langle h\otimes h^{\prime
},\Delta (\hat{g})\,\rangle \,.  \label{du2}
\end{eqnarray}%
Heisenberg doubles provide the associative algebras of quantum phase spaces
endowed with quantum-deformed NC symplectic structure (see e.g. \cite{yu}),
however without the Hopf-algebraic coalgebra sector\footnote{%
Partial coalgebraic structure can be introduced only separately in
generalized coordinate and generalized momenta sectors, described
respectively by dual Hopf algebras. If in quantum phase space we introduce
the bialgebroid coproducts (see Sect. 4), the group coproducts are changed,
but the partial coalgebraic structure in generalized momenta sector can be
preserved.}.

\item[ii)] We introduce the twist-deformed coordinate sector described by $%
\hat{X}_{A}\in \widehat{\mathbb{X}}$, where $\widehat{\mathbb{X}}$ is the $%
\mathbb{H}$-module algebra, however not obtained from $\mathbb{H}$ via
Hopf-algebraic duality\footnote{%
In distinction to the action $h\blacktriangleright g$, the action $%
h\vartriangleright X_{A}$ is described by a white triangle.}. We shall
attach to the quantum phase space algebra $\mathcal{\tilde{H}}^{(10,10)}=(%
\hat{x}_{\mu },\tilde{\Lambda}_{\mu \nu };p_{\mu },M_{\mu \nu })$ the
coalgebraic sector and antipodes, which define the twisted Hopf algebroid
structure (see \cite{xu}, \cite{bp}). For twisted Hopf algebroid $\mathcal{%
\tilde{H}}^{(10,10)}$ the base algebra $(\widehat{\mathbb{X}},\star _{%
\mathcal{F}})$ is endowed with the multiplication defined by the star
product (\ref{star}), which describes the generalized NC coordinates $\hat{X}%
_{A}=$ $(\hat{x}_{\mu },\tilde{\Lambda}_{\mu \nu })$, which will describe
the Lie-algebraic type of noncommutativity.
\end{enumerate}

The plan of our paper is the following: In Sect. 2 following \cite{lmmw} and 
\cite{dlw06}\ we shall describe the pair of dual twist-deformed
Poincare-Hopf algebras $\mathbb{H},\widehat{\mathbb{G}}$. In Sect. 3 we
shall introduce the generalized $D=4$ quantum phase spaces $\mathcal{H}%
^{(10,10)}$ containing space-time translations generators $\hat{\xi}_{\mu }$
as well as $\mathcal{\tilde{H}}^{(10,10)}$ with the quantum Minkowski
space-time coordinates $\hat{x}_{\mu }$. In Sect. 4 by following Xu twisted
bialgebroids framework \cite{xu} we will show how to get for the quantum
phase space $\mathcal{\tilde{H}}^{(10,10)}$ the explicit Hopf algebroid
structure. In Sect. 5 we present final remarks and comments on possible
extensions of presented results.

\section{\protect\bigskip Dual pair of twisted Hopf algebras}

\subsection{\protect\bigskip Twist-deformed Poincare algebra $\mathbb{H}$}

The classical $D=4$ Poincare-Hopf algebra looks as follows%
\begin{align}
\lbrack p_{\mu },p_{\nu }]& =0  \label{pkom} \\
\lbrack M_{\mu \nu },p_{\rho }]& =\eta _{\nu \rho }p_{\mu }-\eta _{\mu \rho
}p_{\nu } \\
\lbrack M_{\mu \nu },M_{\rho \sigma }]& =\eta _{\nu \rho }M_{\mu \sigma
}-\eta _{\mu \rho }M_{\nu \sigma }-\eta _{\nu \sigma }M_{\mu \rho }+\eta
_{\mu \sigma }M_{\nu \rho }  \label{pkom3}
\end{align}%
where $\eta _{\mu \nu }=diag(-1,1,...,1)$ and 
\begin{align}
\Delta _{0}(p_{\mu })=p_{\mu }\otimes 1+1\otimes p_{\mu }& \qquad \Delta
_{0}(M_{\mu \nu })=M_{\mu \nu }\otimes 1+1\otimes M_{\mu \nu }
\label{undef-antip} \\
S_{0}(p_{\mu })=-p_{\mu }& \qquad S_{0}(M_{\mu \nu })=-M_{\mu \nu }
\label{anti} \\
\epsilon _{0}(p_{\mu })=0& \qquad \epsilon _{0}(M_{\mu \nu })=0.
\end{align}

We define twist $\mathcal{F}$ as an element of\thinspace\ $\mathbb{H}\otimes 
\mathbb{H}$ $(\mathbb{H=}\mathcal{U(P)})$ which has an inverse, satisfies
the cocycle condition 
\begin{equation}
\mathcal{F}_{12}\,\left( \Delta _{0}\otimes 1\right) \,\mathcal{F}=\mathcal{F%
}_{23}\,\left( 1\otimes \Delta _{0}\right) \,\mathcal{F}\,  \label{cocy}
\end{equation}%
and the normalization condition 
\begin{equation}
(\epsilon \otimes 1)\mathcal{F}=(1\otimes \epsilon )\mathcal{F}=1\,
\end{equation}%
where $\mathcal{F}_{12}=\mathcal{F}\otimes 1$ and $\mathcal{F}_{23}=1\otimes 
\mathcal{F}$. It is known, that $\mathcal{F}$ does not modify the algebraic
part and the counit, but changes the coproducts and the antipodes in the
following way 
\begin{eqnarray}
\Delta _{\mathcal{F}}(h) &=&\mathcal{F}\circ \,\Delta _{0}(h)\circ \mathcal{F%
}^{-1}\,  \label{dlws2.1} \\
S_{\mathcal{F}}(h) &=&U\,S_{0}(h)\,U^{-1}\,  \label{dlws2.12}
\end{eqnarray}%
where due to (\ref{cocy}) the coproduct $\Delta _{\mathcal{F}}$ is
coassociative, $U=\sum f_{(1)}S(f_{(2)})$ (we use Sweedler's notation $%
\mathcal{F}=\sum f_{(1)}\otimes f_{(2)}$). In such a way we obtain for any
choice of the twist $\mathcal{F}$ the quasitriangular Hopf algebra.\newline

Our family of Abelian twists is \footnote{%
See \cite{dlw06} and \cite{lmmw}; the twists corresponding to special cases $%
u=0$ and $u=\frac{1}{2}$ were considered also respectively in \cite{31ex}
and \cite{32ex}. The twists (\ref{twist-F}) with different $u$ can be
related by coboundary twist (see e.g. \cite{bjt}, Sect. 3.1) which does not
modify the universal $\mathcal{R}$-matrix.} 
\begin{equation}
\begin{split}
\mathcal{F}_{u}& :=\exp (f_{u})=\exp \left( (1-u)\frac{a\cdot p}{2\kappa }%
\otimes \theta ^{\alpha \beta }M_{\alpha \beta }-u\theta ^{\alpha \beta
}M_{\alpha \beta }\otimes \frac{a\cdot p}{2\kappa }\right) \\
& =\exp \left( \frac{1-u}{2}\mathcal{K}^{\alpha \beta }\otimes M_{\alpha
\beta }-\frac{u}{2}M_{\alpha \beta }\otimes \mathcal{K}^{\alpha \beta
}\right)
\end{split}
\label{twist-F}
\end{equation}%
where $a\cdot p=a^{\mu }\eta _{\mu \nu }p^{\nu }$, The parameter $u\in
\lbrack 0,1]$, $\kappa $ is the deformation parameter with dimension of
mass, $a^{2}\equiv a_{\mu }a^{\mu }\in \{-1,0,1\}$, 
\begin{equation}
\mathcal{K}^{\mu \nu }=\frac{a\cdot p}{\kappa }\theta ^{\mu \nu }
\label{mathcalK}
\end{equation}%
and further 
\begin{equation}
\theta ^{\mu \nu }=-\theta ^{\nu \mu },\qquad a_{\mu }\theta ^{\mu \nu }=0.
\end{equation}%
From twists (\ref{twist-F}) one gets the coproducts 
\begin{align}
\Delta _{\mathcal{F}}(p_{\mu })& =p_{\alpha }\otimes (e^{-u\mathcal{K}%
})^{\alpha }{}_{\mu }+(e^{(1-u)\mathcal{K}})^{\alpha }{}_{\mu }\otimes
p_{\alpha }  \label{Delta-p} \\
\Delta _{\mathcal{F}}(M_{\mu \nu })& =M_{\alpha \beta }\otimes (e^{-u%
\mathcal{K}})^{\alpha }{}_{\mu }(e^{-u\mathcal{K}})^{\beta }{}_{\nu
}+(e^{(1-u)\mathcal{K}})^{\alpha }{}_{\mu }(e^{(1-u)\mathcal{K}})^{\beta
}{}_{\nu }\otimes M_{\alpha \beta }  \label{M-Delta} \\
& +\frac{\theta ^{\alpha \beta }}{2\kappa }(a_{\mu }\delta _{\nu }^{\gamma
}-a_{\nu }\delta _{\mu }^{\gamma })\left[ (1-u)p_{\delta }\otimes M_{\alpha
\beta }(e^{-u\mathcal{K}})^{\delta }{}_{\gamma }-uM_{\alpha \beta }(e^{(1-u)%
\mathcal{K}})^{\delta }{}_{\gamma }\otimes p_{\delta }\right]  \notag
\end{align}%
where $\mathcal{K}_{\mu \nu }$ is given by equation \eqref{mathcalK}, and
corresponding antipodes are: 
\begin{align}
S_{\mathcal{F}}(p_{\mu })& =-{(e^{-(1-2u)\mathcal{K}})}^{\alpha }{}_{\mu
}p_{\alpha }  \label{S-p} \\
S_{\mathcal{F}}(M_{\mu \nu })& =-M_{\alpha \beta }{(e^{-(1-2u)\mathcal{K}%
})^{\alpha }{}_{\mu }(e^{-(1-2u)\mathcal{K}})}^{\beta }{}_{\nu }  \label{S-M}
\\
& +\frac{1}{\kappa }(a_{\mu }\delta _{\nu }^{\alpha }-a_{\nu }\delta _{\mu
}^{\alpha })\left[ S(p_{\alpha })+(1-2u)\theta _{\alpha \beta }S(p^{\beta })%
\right]  \notag
\end{align}%
The counit is trivial: 
\begin{equation}
\epsilon (p_{\mu })=0,\qquad \epsilon (M_{\mu \nu })=0.
\end{equation}

\subsection{RTT matrix quantum Poincare group $\widehat{\mathbb{G}}$}

The universal $\mathcal{R}$-matrix ($(a\wedge b=a\otimes b-b\otimes a)$) 
\begin{equation}
\mathcal{R}=\mathcal{F}_{u}^{T}\,\mathcal{F}_{u}^{-1}=\exp [\frac{1}{2}%
(M_{\alpha \beta }\wedge \mathcal{K}^{\alpha \beta })]\,\qquad \qquad
(a\otimes b)^{T}=b\otimes a\,  \label{dlww2.4}
\end{equation}%
can be used for the description of 10-generator deformed $D=4$ Poincar\'{e}
group. Using the $5\times 5$ - matrix realization of the Poincar\'{e}
generators 
\begin{equation}
(M_{\mu \nu })_{\ B}^{A}=\delta _{\ \mu }^{A}\eta _{\nu B}-\delta _{\ \nu
}^{A}\eta _{\mu B}\,\qquad \qquad (P_{\mu })_{\ B}^{A}=-i\delta _{\ \mu
}^{A}\delta _{\ B}^{4}\,  \label{rep}
\end{equation}%
we can show that in (\ref{dlww2.4}) only the linear term is non-vanishing 
\begin{equation}
\mathcal{R}=1\otimes 1+\frac{1}{2}(M_{\alpha \beta }\wedge \mathcal{K}%
^{\alpha \beta }).
\end{equation}

To find the matrix quantum group which is dual to our Hopf algebra $\mathbb{H%
}$ we introduce the following $5\times 5$ - matrices 
\begin{equation}
\mathcal{\hat{T}}_{AB}=\left( 
\begin{array}{cc}
\hat{\Lambda}_{\mu \nu } & \hat{\xi}_{\mu } \\ 
0 & 1%
\end{array}%
\right) \,  \label{dlww2.7}
\end{equation}%
where $\hat{\Lambda}_{\mu \nu }$ parametrizes the quantum Lorentz rotation
and $\hat{\xi}_{\mu }$ denotes quantum translations. In the framework of the 
$\mathcal{FRT}$ procedure \cite{frt}, the algebraic relations defining such
a quantum group $\widehat{\mathbb{G}}$ is described by the following
relation 
\begin{equation}
\mathcal{R}\hat{\mathcal{T}}_{1}\hat{\mathcal{T}}_{2}=\hat{\mathcal{T}}_{2}%
\hat{\mathcal{T}}_{1}\mathcal{R}\,  \label{dlww2.8}
\end{equation}%
while the composition law for the coproduct remains classical 
\begin{equation}
\Delta (\hat{\mathcal{T}}_{AB})=\hat{\mathcal{T}}_{AC}\otimes \hat{\mathcal{T%
}}_{\ B}^{C}\,  \label{dlww2.9}
\end{equation}%
with $\hat{\mathcal{T}}_{1}=\hat{\mathcal{T}}\otimes 1$, $\hat{\mathcal{T}}%
_{2}=1\otimes \hat{\mathcal{T}}$ and quantum $\mathcal{R}$-matrix given in
the representation (\ref{rep}).\newline
In terms of the operator basis $(\hat{\Lambda}_{\mu \nu },\hat{\xi}_{\mu })$
the algebraic relations (\ref{dlww2.8}) describing quantum group algebra can
be written as follows 
\begin{eqnarray}
&&\left[ \hat{\xi}_{\mu },\hat{\xi}_{\nu }\right] =\frac{i}{\kappa }(a_{\nu
}\theta _{\mu }^{\;\alpha }-a_{\mu }\theta _{\nu }^{\;\alpha })\hat{\xi}%
_{\alpha }\,  \label{dlww2.10a} \\
&&[\hat{\xi}_{\mu },\hat{\Lambda}_{\rho \sigma }]=\frac{i}{\kappa }\left( 
\hat{\Lambda}_{\rho \alpha }\theta _{\ \sigma }^{\alpha }a^{\gamma }\hat{%
\Lambda}_{\mu \gamma }-a_{\mu }\theta _{\rho }^{\;\alpha }\hat{\Lambda}%
_{\alpha \sigma }\right) \,  \label{dlww2.10b} \\
&&[\hat{\Lambda}_{\mu \nu },\hat{\Lambda}_{\rho \sigma }]=0\,
\label{dlww2.10c}
\end{eqnarray}%
while the coproduct (\ref{dlww2.9}) takes the well known primitive forms 
\begin{equation}
\Delta \,(\hat{\Lambda}_{\mu \nu })=\hat{\Lambda}_{\mu \rho }\otimes \hat{%
\Lambda}_{\ \nu }^{\rho }\,\qquad \qquad \Delta (\hat{\xi}_{\mu })=\hat{%
\Lambda}_{\mu \nu }\otimes \hat{\xi}^{\nu }+\hat{\xi}_{\mu }\otimes 1\,.
\label{dlww2.11}
\end{equation}%
One can check that coproducts (\ref{dlww2.11}) are consistent with the
commutators (\ref{dlww2.10a})-(\ref{dlww2.10c}).

\bigskip

\subsection{Duality between quantum algebra $\mathbb{H}$ and quantum group $%
\widehat{\mathbb{G}}$}

Two Hopf algebras \ $\mathbb{H},\widehat{\mathbb{G}}$\ are said to be dual
if there exists a nondegenerate bilinear form%
\begin{equation}
\langle \,,\,\rangle \,:\mathbb{H}\mathcal{\times \widehat{\mathbb{G}}%
\longrightarrow }\mathbb{C}\qquad (h,\hat{g})\longrightarrow \langle h,\,%
\hat{g}\,\rangle \,
\end{equation}%
such that the duality relations (\ref{du1})-(\ref{du2}) are satisfied. One
can check, using the following pairing relations%
\begin{equation}
<p_{\mu },\hat{\xi}_{\nu }>\ =-i\eta _{\mu \nu }\quad <M_{\mu \nu },\hat{%
\Lambda}_{\alpha \beta }>\ =\ -(\eta _{\mu \alpha }\eta _{\nu \beta }-\ \eta
_{\nu \alpha }\eta _{\mu \beta })\quad <1,\hat{\Lambda}_{\mu \nu }>\ =\ \eta
_{\mu \nu }  \label{pairing}
\end{equation}%
that we have in our case $(h,h^{\prime }=\{M_{\mu \nu },p_{\mu }\};\,\hat{g},%
\hat{g}^{\prime }=\{\hat{\Lambda}_{\mu \nu },\hat{\xi}_{\mu }\})$%
\begin{eqnarray}
\langle h,\,[\hat{g},\hat{g}^{\prime }]\,\rangle \, &=&\langle \Delta (h),\,%
\hat{g}\otimes \hat{g}^{\prime }-\hat{g}^{\prime }\otimes \hat{g}\,\rangle \,
\label{1dual} \\
\langle \lbrack h,h^{\prime }],\,\hat{g}\rangle \, &=&\langle h\otimes
h^{\prime }-h^{\prime }\otimes h,\Delta (\hat{g})\,\rangle \,.\qquad
\label{2dual}
\end{eqnarray}%
The basic action of $\mathbb{H}$ on $\widehat{\mathbb{G}}$ promoting $%
\widehat{\mathbb{G}}$ to the $\mathbb{H}$-module is given by the relation%
\begin{equation}
h\blacktriangleright \hat{g}=\hat{g}_{(1)}\langle h,\hat{g}_{(2)}\rangle \ .
\end{equation}%
After using (\ref{du1}) one gets the relation%
\begin{eqnarray}
h\blacktriangleright (\hat{g}\hat{g}^{\prime }) &=&\hat{g}\hat{g}%
_{(1)}^{\prime }\langle \Delta h,\hat{g}_{(2)}\otimes \hat{g}_{(2)}^{\prime
}\rangle =\hat{g}\hat{g}_{(1)}^{\prime }\langle h_{(1)},\hat{g}_{(2)}\rangle
\langle h_{(1)},\hat{g}_{(2)}^{\prime }\rangle  \label{covv} \\
&=&(h_{(1)}\blacktriangleright \hat{g})(h_{(2)}\blacktriangleright \hat{g}%
^{\prime })\   \notag
\end{eqnarray}%
what establishes that algebra $\widehat{\mathbb{G}}$ is the $\mathbb{H}$%
-module.

\section{Heisenberg double $\mathcal{HD}$ and generalized $D=4$ quantum
phase spaces}

\bigskip Using the formula (\ref{smash}), in Heisenberg double framework we
can obtain cross commutators between the algebra $\mathbb{H}$ and group $%
\widehat{\mathbb{G}}$ by the following relation%
\begin{equation}
\lbrack h,\hat{g}]=\hat{g}_{(2)}\langle h_{(1)},\hat{g}_{(1)}\,\rangle
\,h_{(2)}-\hat{g}h\qquad h=\{M_{\mu \nu },p_{\mu }\};\,\hat{g}=\{\hat{\Lambda%
}_{\mu \nu },\hat{\xi}_{\mu }\}.\qquad  \label{HD}
\end{equation}

Using pairing (\ref{pairing}) and coproducts (\ref{Delta-p}), (\ref{M-Delta}%
) and (\ref{dlww2.11}) we get%
\begin{eqnarray}
\lbrack p_{\mu },\hat{\xi}_{\nu }] &=&-i(e^{-u\mathcal{K}})_{\nu \mu }+i(1-u)%
\frac{a_{\nu }}{\kappa }{\theta _{\mu }}^{\alpha }p_{\alpha }  \label{defo_1}
\\
\lbrack p_{\mu },\hat{\Lambda}_{\rho \sigma }] &=&0  \label{defo2} \\
\lbrack M_{\mu \nu },\hat{\Lambda}_{\rho \sigma }] &=&\frac{u}{\kappa }%
\theta _{\rho }^{\;\alpha }\hat{\Lambda}_{\alpha \sigma }(a_{\mu }p_{\nu
}-a_{\nu }p_{\mu })  \label{defo_3} \\
&&+\hat{\Lambda}_{\alpha \sigma }(e^{-u\mathcal{K}})_{\;[\mu }^{\alpha
}(e^{-u\mathcal{K}})_{\rho \nu ]}  \notag \\
\lbrack M_{\mu \nu },\hat{\xi}_{\rho }] &=&\hat{\xi}_{\gamma }(e^{u\mathcal{K%
}})_{[\mu }^{\;\;\gamma }(e^{u\mathcal{K}})_{\nu ]\rho }+\frac{u}{\kappa }%
\theta _{\rho }^{\;\alpha }\hat{\xi}_{\alpha }(a_{\mu }p_{\nu }-a_{\nu
}p_{\mu })  \label{defo_4} \\
&&+\frac{i(u-1)}{2\kappa }\theta ^{\alpha \beta }M_{\alpha \beta }a_{[\mu
}(e^{-u\mathcal{K}})_{\rho \upsilon ]}+\frac{i(u-1)}{\kappa }a_{\rho
}M_{[\mu \alpha }\theta _{\ \nu ]}^{\alpha }  \notag
\end{eqnarray}

\bigskip One can show that the generators of quantum Poincare group algebra,
satisfying relations (\ref{dlww2.10a})-(\ref{dlww2.10c}), (\ref{defo_1})-(%
\ref{defo_4}) can be expressed in terms of classical group parameters $(\xi
_{\mu },\Lambda _{\rho \sigma })$ as follows 
\begin{eqnarray}
\hat{\xi}_{\mu } &=&\xi _{\alpha }(e^{-u\mathcal{K}})_{\mu }^{\;\alpha
}+(1-u)\frac{ia_{\mu }}{2\kappa }\theta M \\
\hat{\Lambda}_{\rho \sigma } &=&(e^{-u\mathcal{K}})_{\rho }^{\;\alpha
}\Lambda _{\alpha \beta }(e^{-\frac{1}{\kappa }(p\Lambda a)\theta
})_{\,\;\sigma }^{\beta }.
\end{eqnarray}

The group manifold $\widehat{\mathbb{G}}$ as $\mathbb{H}$-module algebra can
be promoted to the base algebra of Hopf algebroid. We get in such a way the
application of Lu construction \cite{lu}, which provides for Heisenberg
double $\mathcal{H}\mathcal{D}=\mathbb{H}\rtimes \widehat{\mathbb{G}}$, with
generators $(\hat{\xi}_{\mu },\hat{\Lambda}_{\mu \nu };p_{\mu },M_{\mu \nu
}) $ the Hopf algebroid structure\footnote{%
See \cite{lu}, Sect. 5. The compact formula for the corresponding target map
is under consideration. We add that in \cite{lu} it was proved explicitly
that the Heisenberg doubles of finite-dimensional Hopf algebras carry a Hopf
algebroid structure. Our Hopf algebra $H$ is infinite-dimensional, however
filtered by finite-dimensional subalgebras $H_{N\text{,}}$ and the limit $%
N\rightarrow \infty $ should be considered for delivering the rigorous proof.%
}.

\bigskip Let us introduce in place of relations (\ref{dlww2.10b}) the
following set of parameter-dependent algebraic relations ($s-$real parameter)%
\begin{equation}
\lbrack \hat{\xi}_{\mu }^{(s)},\hat{\Lambda}_{\rho \sigma }^{(s)}]=\frac{i}{%
\kappa }\left( s\hat{\Lambda}_{\rho \alpha }^{(s)}\theta _{\ \sigma
}^{\alpha }a^{\gamma }\hat{\Lambda}_{\mu \gamma }^{(s)}-a_{\mu }\theta
_{\rho }^{\;\alpha }\hat{\Lambda}_{\alpha \sigma }^{(s)}\right) \,.
\label{gengen}
\end{equation}%
One can show that the relation (\ref{gengen}) forms together with relations (%
\ref{dlww2.10a}), (\ref{dlww2.10c}) the consistent algebra of associative
generalized quantum Poincare phase spaces, satisfying Jacobi relations.
However, for $\xi =0$ and $\xi =1$ one can supplement the Hopf algebroid
structure, namely

\begin{enumerate}
\item[i)] if $s=1$ $(\hat{\xi}_{\mu }^{(1)}\equiv \hat{\xi}_{\mu },\hat{%
\Lambda}_{\rho \sigma }^{(1)}\equiv \hat{\Lambda}_{\rho \sigma })$ one can
construct the Hopf algebroid $\mathcal{H}^{(10,10)}$ with base algebra $%
\widehat{\mathbb{G}}=(\hat{\xi}_{\mu },\hat{\Lambda}_{\mu \nu })$ described
by algebraic quantum Poincare group manifold.

\item[ii)] if $s=0$ $(\hat{\xi}_{\mu }^{(0)}\equiv \hat{x}_{\mu },\hat{%
\Lambda}_{\rho \sigma }^{(0)}\equiv \tilde{\Lambda}_{\rho \sigma })$ one
gets alternative Hopf algebroid, denoted by $\mathcal{\tilde{H}}^{(10,10)}$,
with algebra $\mathbb{\hat{X}=}(\hat{x}_{\mu },\tilde{\Lambda}_{\mu \nu })$
as base algebra, with the multiplication given by the star product formula (%
\ref{star}). The formula (\ref{star}) can be also written as follows%
\begin{equation}
f(X)\star _{\mathcal{F}}k(X^{\prime })=\widehat{f(X)}\vartriangleright
k(X^{\prime })  \label{50e}
\end{equation}%
where $\widehat{f(X)}$ denotes the noncommutative representation of $f(\hat{X%
})$ defined as follows%
\begin{equation}
f(\hat{X})\simeq \widehat{f(X)}=m[\mathcal{F}^{-1}(\vartriangleright \otimes
1)(f(X)\otimes 1)]  \label{51}
\end{equation}%
where in formulae (\ref{50e}), (\ref{51}) we use (see also (\ref{star}))%
\begin{equation}
h\vartriangleright X_{A}=[h,X_{A}],\;\qquad h=\{p_{\mu },M_{\mu \nu
}\},X_{A}=\{x_{\mu },\Lambda _{\mu \nu }\}
\end{equation}%
and for Lorentz sector we obtain 
\begin{equation}
\lbrack M_{\mu \nu },\Lambda _{\rho \sigma }]=\eta _{\rho \nu }\Lambda _{\mu
\sigma }-\eta _{\rho \mu }\Lambda _{\nu \sigma }.  \label{nondefM}
\end{equation}%
Substituting in (\ref{51}) the twist (\ref{twist-F}) we get the following
explicit formulas describing base algebra $\hat{X}_{A}=\{\hat{x}_{\mu },%
\tilde{\Lambda}_{\mu \nu }\}$%
\begin{eqnarray}
\hat{x}_{\mu } &=&m[\mathcal{F}^{-1}(\vartriangleright \otimes 1)(x_{\mu
}\otimes 1)]  \label{dd1} \\
&=&x_{\alpha }{(e^{-u\mathcal{K}})_{\mu }}^{\alpha }+(1-u)\frac{ia_{\mu }}{%
2\kappa }\theta M  \notag \\
\tilde{\Lambda}_{\rho \sigma } &=&m[\mathcal{F}^{-1}(\vartriangleright
\otimes 1)(\Lambda _{\rho \sigma }\otimes 1)]  \label{dd2} \\
&=&{(e^{-u\mathcal{K}})_{\rho }}^{\alpha }\Lambda _{\alpha \sigma }  \notag
\end{eqnarray}%
satisfying the following algebraic relations 
\begin{eqnarray}
&&\left[ \hat{x}_{\mu },\hat{x}_{\nu }\right] =\frac{i}{\kappa }(a_{\nu
}\theta _{\mu }^{\;\alpha }-a_{\mu }\theta _{\nu }^{\;\alpha })\hat{x}%
_{\alpha }\,  \label{a1} \\
&&[\hat{x}_{\mu },\tilde{\Lambda}_{\rho \sigma }]=-\frac{i}{\kappa }a_{\mu
}\theta _{\rho }^{\;\alpha }\tilde{\Lambda}_{\alpha \sigma }\,  \label{a2} \\
&&[\tilde{\Lambda}_{\mu \nu },\tilde{\Lambda}_{\rho \sigma }]=0\,.
\label{a3}
\end{eqnarray}

The generators $\Lambda _{\rho \sigma }$ and $M_{\mu \nu }$ satisfying the
relation (\ref{nondefM}) describe the pair of undeformed dual variables in
Lorentz sector, defined by the limit $\kappa \rightarrow \infty $ of the
formula (\ref{defo_3}).
\end{enumerate}

\section{Hopf algebroid structures}

The Hopf algebroid is described by the data $\mathcal{B}(H,A;s,t;\tilde{%
\Delta},S,\epsilon )$ where $H$ is total algebra, $A$ the base algebra, and
bialgebroidal coproducts $\tilde{\Delta}$ are described by the maps $%
H\rightarrow H\otimes _{A}H$ into $(A,A)$ bimodules $H\otimes _{A}H$, which
do not inherit the $H$ algebra structure. The bimodule $(A,A)$ property
follows from the assumed existence of two maps:

\begin{enumerate}
\item[i)] source map $s(a):A\rightarrow H$ (homomorphism),

\item[ii)] target maps $t(a):A\rightarrow H$ (antihomomorphism),
\end{enumerate}

where 
\begin{equation}
\lbrack s(a),t(b)]=0\qquad a,b\in A\qquad s(a),t(b)\in H.  \label{image}
\end{equation}%
The relation (\ref{image}) permits to introduce the basic $(A,A)$ bimodule
formula, namely $ahb=ht(a)s(b)$. Further, the algebra $H\otimes _{A}H$ can
be defined as the quotient of \ $H\otimes H$ by left ideal $\mathcal{I}_{L}$
generated by the following elements$\footnote{%
The use of left ideal (\ref{ideal1})-(\ref{ideal2}) describes right
bialgebroid $\mathcal{\tilde{H}}^{R}$. The elements of left bialgebroid $%
\mathcal{\tilde{H}}^{L}$ are intertwined with $\mathcal{\tilde{H}}^{R}$ by
antipode map $S.$}$%
\begin{equation}
\mathcal{I}_{L}=s(a)\otimes 1-1\otimes t(a)\qquad a\in A.  \label{ideal1}
\end{equation}%
The canonical choice $s(a)=a$ of the source map leads to canonical form of (%
\ref{ideal1})%
\begin{equation}
\mathcal{I}_{L}^{c}=a\otimes 1-1\otimes t^{c}(a)\qquad a\in A  \label{ideal2}
\end{equation}%
with the target map $t^{c}(a)$ determining the explicit form of coalgebra
gauge freedom (see e.g. \cite{lws}). Further the canonical coproducts of the
elements of base algebra $A$ are chosen to be half-primitive (see e.g. \cite%
{lu}-\cite{xu})%
\begin{equation}
\tilde{\Delta}(\hat{a})=\hat{a}\otimes 1\,\qquad \qquad \hat{a}\in A
\label{bacop}
\end{equation}%
i.e. the coproducts (\ref{bacop}) are homomorphic in trivial way to the
algebra structure of $A$.

In this paper we have two choices of algebra $A$: given by quantum group
algebra $\widehat{\mathbb{G}}$, with quantum space-times translations $\hat{%
\xi}_{\mu }$, and by generalized coordinates sector $\widehat{\mathbb{X}}$
with quantum-deformed Minkowski space-time coordinates introduced as the
module algebra of twisted Poincare algebra. Below we shall consider in some
detail the second choice.

If the quantum deformation is generated by twist, the bialgebroid source and
target map are described by the formulae analogous to relations (\ref{51})
(see \cite{xu},\cite{bp}). For the choice (\ref{twist-F}) of twist factor
one gets:

\begin{itemize}
\item source map%
\begin{eqnarray}
s_{\mathcal{F}}(\hat{x}_{\mu }) &=&m[\mathcal{F}^{-1}(\vartriangleright
\otimes 1)(s_{0}(x_{\mu })\otimes 1)] \\
&=&x_{\alpha }{(e^{-u\mathcal{K}})_{\mu }}^{\alpha }+(1-u)\frac{ia_{\mu }}{%
2\kappa }\theta M\equiv \hat{x}_{\mu }  \notag \\
s_{\mathcal{F}}(\tilde{\Lambda}_{\rho \sigma }) &=&m[\mathcal{F}%
^{-1}(\vartriangleright \otimes 1)(s_{0}(\Lambda _{\rho \sigma })\otimes 1)]
\\
&=&{(e^{-u\mathcal{K}})_{\rho }}^{\alpha }\Lambda _{\alpha \sigma }\equiv 
\tilde{\Lambda}_{\rho \sigma }  \notag
\end{eqnarray}

\item target map%
\begin{eqnarray}
t(\hat{x}_{\mu }) &=&m[(\mathcal{F}^{-1})^{\tau }(\vartriangleright \otimes
1)(t_{0}(x_{\mu })\otimes 1)] \\
&=&x_{\alpha }{(e^{(1-u)\mathcal{K}})_{\mu }}^{\alpha }-u\frac{ia_{\mu }}{%
2\kappa }\theta M=\hat{x}_{\alpha }{(e^{\mathcal{K}})_{\mu }}^{\alpha }-i%
\frac{a_{\mu }}{2\kappa }\theta M  \notag \\
t(\tilde{\Lambda}_{\rho \sigma }) &=&m[(\mathcal{F}^{-1})^{\tau
}(\vartriangleright \otimes 1)(t_{0}(\Lambda _{\rho \sigma })\otimes 1)] \\
&=&{(e^{(1-u)\mathcal{K}})_{\rho }}^{\alpha }\Lambda _{\alpha \sigma }={(e^{%
\mathcal{K}})_{\rho }}^{\alpha }\tilde{\Lambda}_{\alpha \sigma }.  \notag
\end{eqnarray}
\end{itemize}

One can check that images of source and target maps commute as follows%
\begin{eqnarray}
\lbrack s(\hat{X}_{A}),s(\hat{X}_{B})] &=&\mathcal{C}{_{AB}}^{C}s(\hat{X}%
_{C})  \label{alg1} \\
\lbrack t(\hat{X}_{A}),t(\hat{X}_{B})] &=&-\mathcal{C}{_{AB}}^{C}t(\hat{X}%
_{C})  \label{alg2} \\
\lbrack s(\hat{X}_{A}),t(\hat{X}_{B})] &=&0  \label{alg3}
\end{eqnarray}%
where constant structures $\mathcal{C}{_{AB}}^{C}$ describe the Lie
algebraic structure of commutation relations (\ref{a1})-(\ref{a3}). The
coproducts of base algebra elements are given by relations (\ref{bacop}),
and the coalgebra of generalized momentum sector $H$ is described by
Hopf-algebraic formulae (\ref{dlws2.1}) and (\ref{Delta-p})-(\ref{M-Delta}).

In the case of our twisted Hopf algebroid the canonical ideal (\ref{ideal2})
is the following%
\begin{eqnarray}
\mathcal{I}_{L}^{c}(\hat{x}_{\mu }) &=&\hat{x}_{\mu }\otimes 1-1\otimes (%
\hat{x}_{\alpha }{(e^{\mathcal{K}})_{\mu }}^{\alpha }-i\frac{a_{\mu }}{%
2\kappa }\theta M) \\
\mathcal{I}_{L}^{c}(\tilde{\Lambda}_{\rho \sigma }) &=&\tilde{\Lambda}_{\rho
\sigma }\otimes 1-1\otimes {(e^{\mathcal{K}})_{\rho }}^{\alpha }\tilde{%
\Lambda}_{\alpha \sigma }
\end{eqnarray}%
with counits given by the canonical formula%
\begin{equation}
\epsilon (\widehat{X}_{A})=m[\mathcal{F}^{-1}(\triangleright \otimes
1)(\epsilon _{0}(X_{A})\otimes 1)]=\widehat{X}_{A}  \label{jed}
\end{equation}%
and $\epsilon (h)=0$ and $\epsilon (1)=1$. The antipodes should satisfy the
following relations%
\begin{eqnarray}
S(t(\widehat{X}_{A})) &=&s(\widehat{X}_{A})=\widehat{X}_{A}  \label{ant1} \\
m[(1\otimes S)\circ \tilde{\Delta}] &=&s\epsilon =\epsilon  \label{antt} \\
m[(S\otimes 1)\circ \tilde{\Delta}] &=&t\epsilon S.  \label{ant2}
\end{eqnarray}%
Using (\ref{ant1}) one gets explicit formulae for the antipodes%
\begin{eqnarray}
S(\widehat{x}_{\mu }) &=&{(e^{\mathcal{K}})_{\mu }}^{\alpha }\widehat{x}%
_{\alpha }-i\frac{a_{\mu }}{2\kappa }\theta ^{\alpha \beta }M_{\alpha \beta
}=t(\widehat{x}_{\mu }) \\
S(\tilde{\Lambda}_{\mu \nu }) &=&{(e^{\mathcal{K}})_{\mu }}^{\alpha }\tilde{%
\Lambda}_{\alpha \nu }=t(\tilde{\Lambda}_{\mu \nu }).
\end{eqnarray}%
In our case we have $S^{2}=1$ and the anchor map is not needed (see \cite{lu}%
-\cite{brzez}).

\section{Final remarks}

The generalization of Hopf algebras, which we define over commutative ring,
are generalized to Hopf-algebraic structures over noncommutative ring,
called base algebra $A$, leads to the notion of Hopf algebroids. If the
algebra sector of Hopf algebroid $\mathcal{B}$ is endowed with
(pre)symplectic structure, as it occurs in Heisenberg double construction
presented here, we obtain various models of quantum-deformed phase space
algebras. We presented in this paper two types of quantum-deformed phase
spaces with Hopf algebroid structure: with coordinates described by
quantum-deformed Poincare group algebra and with NC space-time coordinates
generated by twist-dependent star product formula (\ref{50e}).

Our considerations have been illustrated by explicit example of Drinfeld
twist, but our considerations can be generalized to arbitrary Drinfeld twist
case. It has been shown (see \cite{brzez}, Sect. 4) that there exists a
generic construction of bialgebroid associated with $\mathcal{FRT}$
quantization method of quantum matrix groups, which also in general case
provides the braid-commutative choice of base algebras defining the
Yetter-Drinfeld module \cite{brzez}.

Hopf algebroids are defined in their coalgebraic sectors by the coproducts $%
\tilde{\Delta}:H\rightarrow H\otimes H$ defined modulo the coproduct gauge
freedom (see e.g. \cite{lws}; we stress that $\otimes $ denotes standard
tensor product). It is an interesting problem to characterize the algebraic
manifold $\mathcal{C}$ parametrizing the coproduct gauge transformations,
and subsequently define the coproduct gauge-invariant quantum two-particle
phase space as the factor algebra $H\otimes H/\mathcal{C}$ which does not
depend on the choice of coproduct gauge.

We add that $\mathcal{HD}$ construction and corresponding Hopf algebroid
structures can be introduced also for infinite-dimensional affine algebras,
which are linked with integrable many-body systems of Cologero type (e.g.
Cologero-Moser \cite{fr1}-\cite{fr2} and Ruijsenaars-Schneider \cite{fr3}-%
\cite{fr4} integrable hierarchies) as well as applied for dynamical quantum
groups, described by parameter-dependent (dynamical) Yang-Baxter equations
(see e.g. \cite{39pap}). Other extension, which has as well the link with
algebroid structures, is related with recent development in string models
with applications of generalized geometries and T-duality covariance. Such
generalized phase space dynamics formulated in the framework of string
theories, related with double field theories, has been recently investigated
under the name of metastring theories (see e.g. \cite{fr5}, \cite{frs}).

\section*{Acknowledgements}

One of the authors (J.L.) would like to thank Andrzej Borowiec and Zoran
Skoda for valuable comments. J.L. and M.W. have been suported by Polish
National Science Center, project 2017/27/B/ST2/01902.

\end{document}